\documentclass[aps, prl, 10pt, twocolumn,superscriptaddress,longbibliograph]{revtex4-2}
\usepackage{
    physics,
    mathtools,
    amssymb,
    bm,
    xcolor,
    graphicx,
}
\usepackage[bookmarks=false,linkcolor=blue,urlcolor=blue,colorlinks,citecolor=blue]{hyperref}
\usepackage[capitalize]{cleveref}
\Crefname{figure}{Fig.}{Figs.}
\usepackage[utf8]{inputenc}
\usepackage{epsfig}
\usepackage{graphicx}
\usepackage{float}
\usepackage{dcolumn}
\usepackage{natbib}
\usepackage{xcolor}
\usepackage{bm}
\usepackage[normalem]{ulem}
\usepackage{amsmath,bbm}

\DeclareUnicodeCharacter{2212}{-}

\newcommand{\ceq}[1] {(\ref{#1})}

\begin{document}

\title{Enhanced superconducting diode effect in hybrid Josephson junctions}

\author{Peng Yu}
\affiliation{Center for Quantum Information Physics, New York University, New York, NY 10003, USA}
\author{Han Fu}
\affiliation{Department of Physics, William \& Mary, Williamsburg, Virginia 23187, USA} 
\affiliation{Department of Physics, Florida Atlantic University, Boca Raton, Florida, 33431, USA}
\author{William F. Schiela}
\affiliation{Center for Quantum Information Physics, New York University, New York, NY 10003, USA}
\author{William Strickland}
\affiliation{Center for Quantum Information Physics, New York University, New York, NY 10003, USA}
\author{Bassel Heiba Elfeky}
\affiliation{Center for Quantum Information Physics, New York University, New York, NY 10003, USA}
\author{S. M. Farzaneh}
\affiliation{Center for Quantum Information Physics, New York University, New York, NY 10003, USA}
\author{Jacob Issokson}
\affiliation{Center for Quantum Information Physics, New York University, New York, NY 10003, USA}
\author{{Wei Pan}}
\affiliation{Sandia National Laboratories, Albuquerque, New Mexico 87185, USA}
\author{Enrico Rossi}
\affiliation{Department of Physics, William \& Mary, Williamsburg, Virginia 23187, USA} 
\author{Javad Shabani}
\affiliation{Center for Quantum Information Physics, New York University, New York, NY 10003, USA}
\date{\today}

\begin{abstract}
The superconducting diode effect (SDE) has recently been observed in various systems, sparking interest in novel superconducting devices and offering a new platform to probe intrinsic material properties. Josephson junctions with strong Rashba spin-orbit coupling have exhibited nonreciprocal critical currents under applied magnetic fields. In this work, we investigate the SDE in Josephson junctions incorporating periodic hole arrays patterned into the superconducting leads on InAs heterostructures with epitaxial aluminum. We observe an enhanced diode effect when a top gate depletes the 2DEG in the region of the hole arrays, while preserving the overall supercurrent. Theoretical analysis shows that the physics behind this phenomenon is the increased difference of transparency between different bands in the junction.
These results highlight a new pathway for engineering and controlling nonreciprocal superconducting transport in hybrid systems.

\end{abstract}

\maketitle
The superconducting diode effect (SDE) has recently sparked widespread interest in both experimental and theoretical condensed matter research, owing to its promising potential in advanced superconducting technologies \cite{Ando2020, Yuan2022, Misaki2021, Davydova2022, Gutfreund2023, Bauriedl2022, Wu2022, Pal2022, Gupta2023, Daido2022,cuozzo2024,yu2024z,travaglini2025}. Analogous to conventional semiconductor diodes—essential components in modern electronics—superconducting diodes are envisioned as key elements in next-generation low-dissipation circuits. However, unlike semiconductor diodes, which suppress current in one direction, the SDE manifests as a nonreciprocal critical current, i.e., a directional asymmetry in the maximum supercurrent the system can carry. Among various systems that can host the SDE, recent studies have demonstrated the SDE in Josephson junctions (JJ) incorporating  Rashba spin-orbit coupling (SOC) and broken time-reversal symmetry \cite{Yuan2022, Chen2018, Baumgartner2022, Baumgartner2022_spin-orbit, Turini2022, cuozzo2024}, and even in systems lacking explicit SOC \cite{Davydova2022, Banerjee2023, Hou2023,yu2024z}. In such setups, the SDE is frequently attributed to finite-momentum pairing mechanisms, making it a sensitive probe of the underlying material properties \cite{Daido2022prb, Ilic2022, Ilic2024, Pal2022, Turini2022}. Despite growing experimental evidence, realizing a strong and tunable SDE remains a critical challenge for practical applications. Among potential platforms, Josephson junctions are particularly appealing due to their enhanced electrostatic tunability \cite{Yu2025}, especially when compared to high-carrier-density systems such as metallic or van der Waals materials \cite{Margineda2023, Kim2024}. Theoretical proposals suggest that both the geometry and the microscopic properties of Josephson junctions can significantly influence the SDE \cite{Davydova2022, Costa2023, Wang2025}.  

In this work, we investigate the SDE in planar Josephson junctions fabricated from materials with strong spin-orbit coupling \cite{Farazaneh2024}. Building on our previous findings, where the diode effect was shown to depend on superconducting contact width and the orientation of an in-plane magnetic field \cite{Lotfizadeh2024}, we now explore how the SDE can be tuned via various electrostatic gating in a newly designed junction geometry. Our results reveal an unexpected and unconventional enhancement of SDE efficiency that cannot be simply explained by orbit effect \cite{Banerjee2023, schiela2025}, changes in chemical potential or superconducting coupling strength in the junction. Instead, from numerical simulations and analysis, we find that the observed behavior is linked to modifications in Andreev reflection processes, suggesting a novel mechanism for controlling the superconducting diode effect. 
\begin{figure*} \includegraphics[width=1\textwidth]{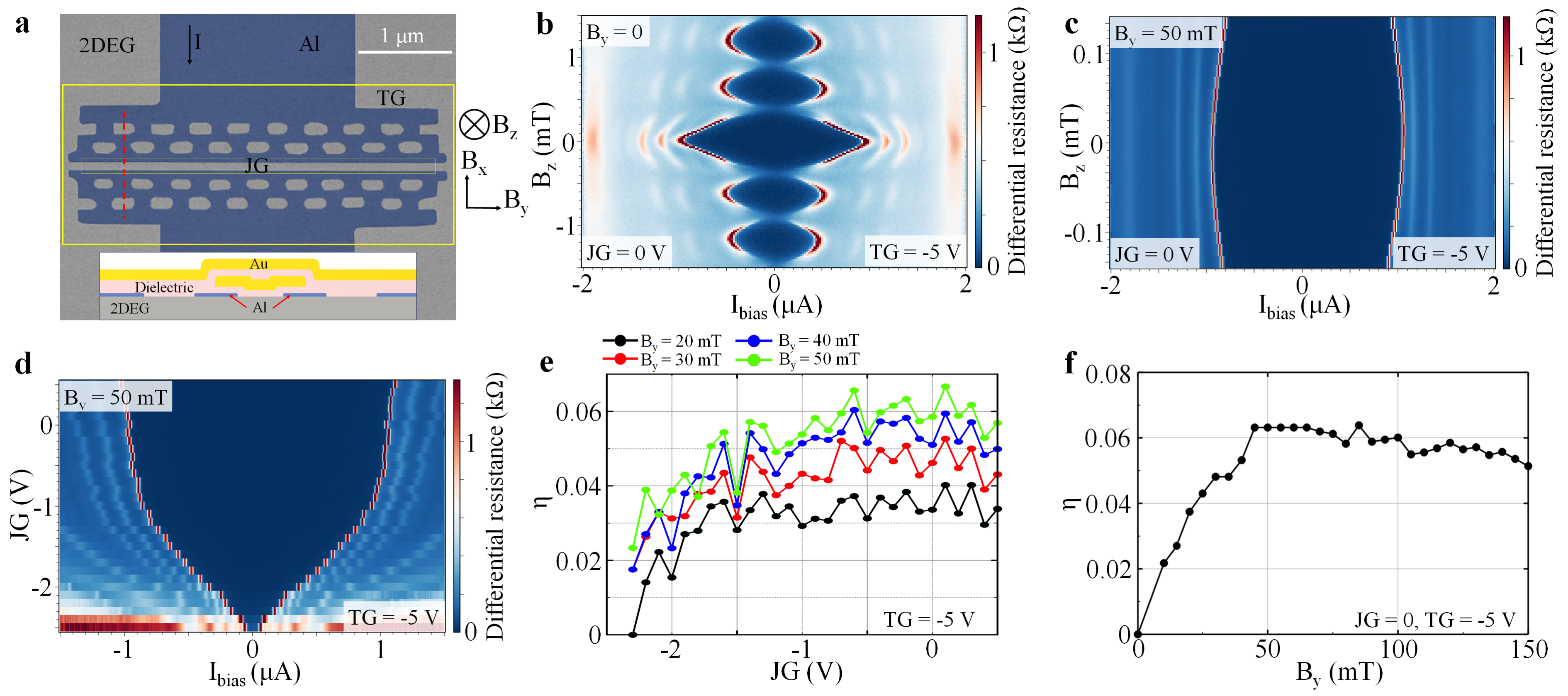}
  \caption{(a) False-color scanning electron micrograph of the measured device and schematic of the device and the material stacks. (b) Differential resistance as a function of the bias current and $B_z$. (c) Differential resistance as a function of the bias current and TG voltages at $B_z$ = 0 and $B_y$ = 50 mT. (d) Differential resistance as a function of the bias current and JG voltages at $B_z$ = 0  and $B_y$ = 50 mT. (e) Diode effect efficiency $\eta$ as a function of JG voltages. (f) Diode effect efficiency $\eta$ as a function of $B_y$. }
 \label{fig1}
\end{figure*}

The Josephson junction devices investigated in this work are fabricated on InAs heterostructures with epitaxial superconducting Al thin films serving as the superconducting layer. Device fabrication involves a two-step process: first, a deep etch is performed to define the mesa structure, followed by a selective etching of the Al layer to pattern the Josephson junction geometry. The junction weak link is 4 $\mu$m long with a width of around 100 nm. Two rows of periodic holes are intentionally etched on each side of the Al contacts as shown in Fig.1(a).  Electrostatic control of the device is achieved using a double-gate architecture. The first layer consists of a junction gate (JG), which is positioned directly over the junction region to modulate the chemical potential in the central section of the junction. A second dielectric layer electrically isolates the JG from the top gate (TG), which constitutes the second gate layer. The TG spans a significantly larger area of the device, including the regions containing the etched holes, thereby allowing independent tuning of the surrounding electronic environment. A schematic illustration of the device geometry and material stack is shown in the inset of Fig.1(a). The JG is designed to be shorter than the junction by 100 nm at each end. Although JG itself cannot fully deplete the junction (see SM), since it does not cover the ends of the junction, the chemical potential in the whole junction can be fully controlled by using JG and TG together (see SM). In particular, TG plays two roles: depletion of the ends of the junction and depletion of the 2DEG in the region of the etched holes on the Al contacts. Based on the results of various gate scans (see SM), we expect the TG to deplete the region of the holes when the voltage is below - 2.5 V. As discussed in Ref.~\onlinecite{Yu2025}, the depletion of the 2DEG in the region of the holes has a nontrivial effect on the supercurrent. A significant supercurrent enhancement has been observed when the holes are depleted at finite magnetic fields, which we attribute to a strengthened Andreev reflection induced by the TG \cite{Yu2025}. In this manuscript, we focus on the SDE observed at low magnetic fields. All measurements in this study were performed at around 30 mK in a dilution refrigerator equipped with a three-axis vector magnet. As shown in Fig.~\ref{fig1}~(a), the magnetic field is defined with respect to the sample geometry such that the z-axis is oriented perpendicular to the sample plane, while the x- and y-axes lie in-plane, aligned parallel and perpendicular to the current flow, respectively. We used standard low-frequency lock-in amplification techniques with excitation currents of no more than 10 nA and frequencies of 17 and 77 Hz. The thin Al film used in our devices is highly sensitive to out-of-plane magnetic fields, making precise field alignment essential for reliable measurements. To ensure accurate alignment of a desired in-plane magnetic field —e.g., $B_y$ = 50 mT, we perform an out-of-plane field ($B_z$) scan while holding $B_y$ = 50 mT fixed. The actual $B_z$ = 0 condition is determined by identifying the center of the Fraunhofer interference pattern, which corresponds to zero net flux through the junction. Once $B_z$ is aligned to this true zero, we fix the field configuration and perform bias current versus gate voltage scans to investigate the nonreciprocity of the critical current (see SM for more details about the measurement).  

\begin{figure*} \includegraphics[width=1\textwidth]{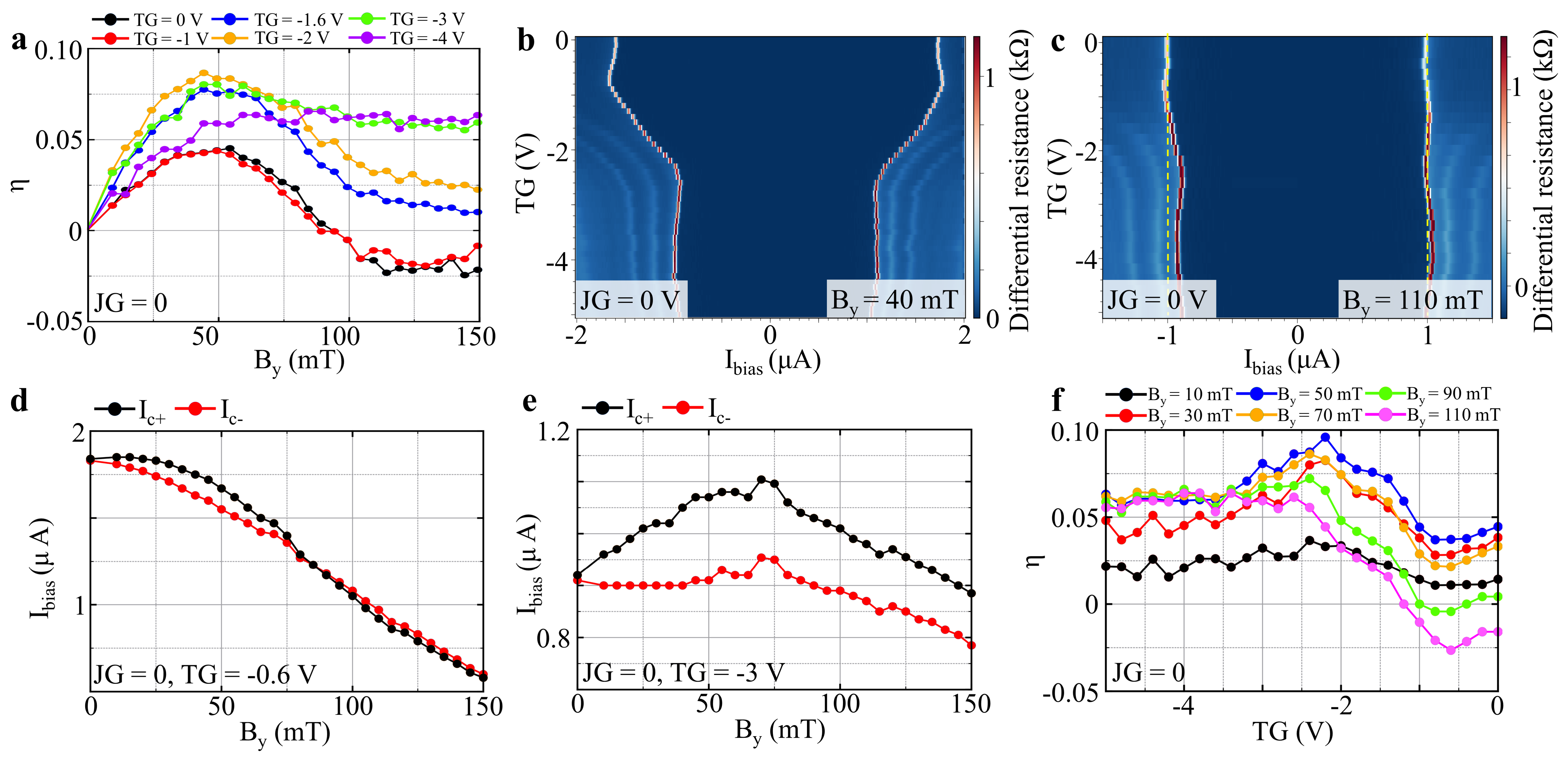}  
  \caption{ (a) The diode effect efficiency $\eta$ as a function of $B_y$ for different TG voltages at JG = 0. (b) Differential resistance as a function of the bias current and TG voltages at $B_y$ = 40 mT. (c) Differential resistance as a function of the bias current and TG voltages at $B_y$ = 110 mT. (d) Absolute values of the critical currents as a function of $B_y$ for JG = 0 and TG = -0.6 V. (e) Absolute values of the critical currents as a function of $B_y$ for JG = 0 and TG = -3 V. (f) The diode effect efficiency $\eta$ as a function of TG voltages for different $B_y$ at JG = 0. 
 }
 \label{fig2}
\end{figure*}
In Fig.~\ref{fig1}~(b) and (c), we sweep $B_z$ while JG is set to 0 and TG is set to -5 V. The choice of TG of -5 V is to suppress SQUID-like interference effects arising from parallel conduction paths at the ends (see SM).
In these figures we present the differential resistance as a function of the bias current and the out-of-plane field $B_z$ for $B_y$ = 0 and $B_y$ = 50 mT. We observe that, at zero magnetic field, the critical current is consistently larger when the bias current is swept from zero to high values compared to vice versa. This asymmetry is likely attributed to heating effects (see SM for a detailed discussion), which introduce a hysteresis in the current-voltage characteristics. To mitigate artifacts arising from this effect, all measurements presented in this work were performed using a bias sweep from positive bias to zero and then from negative bias to zero, unless explicitly stated otherwise. The reason we choose to sweep the bias from high bias to zero is that the cold electron branch usually exhibits a broad switching distribution near the Fraunhofer maximum \cite{Lotfizadeh2024}. We therefore only derive the values of the critical current from the hot electron branch. Since we focus on the diode effect, using the hot electron branch will not compromise the robustness of our conclusions (see SM for a detailed discussion). At zero in-plane field (Fig.~\ref{fig1}~(b)), the scan reveals a Fraunhofer pattern that has equal critical currents measured at positive and negative bias within the range of measurement error, indicating the absence of SDE under these conditions. We have also verified that there is no SDE when the in-plane field is parallel to the supercurrent (see SM). We repeat the $B_z$ sweep at $B_y$ = 50 mT for a reduced $B_z$ range and focused on the main lobe, as shown in Fig.~\ref{fig1}~(c). A clear asymmetry emerges between the positive and negative critical currents across the entire range of $B_z$. This indicates a superconducting diode effect under an in-plane magnetic field perpendicular to the supercurrent. In Fig.~\ref{fig1}~(d), we plot critical current as a function of bias current and JG voltages at TG = -5 V and $B_z$ = 0 and $B_y$ = 50 mT. The supercurrent can be fully depleted by the JG and a clear SDE can be observed for the entire range of JG voltages. 

The diode effect's efficiency $\eta$ ($\eta=\frac{I_{c+} - |I_{c-}|}{I_{c+} + |I_{c-}|}$) characterizes the strength of the SDE. We extract $\eta$ from bias current versus gate scans, and investigate its dependence on gate voltages and in-plane field $B_y$. In Fig.~\ref{fig1}~(e), $\eta$ is plotted as a function of the JG voltages for various values of $B_y$, with the top gate fixed at TG = −5 V. As can be seen, $\eta$ generally decreases as voltages on the JG decrease. In contrast, increasing $B_y$ from 20 mT to 50 mT leads to a clear enhancement of $\eta$. These trends are consistent with expectations: lowering the JG voltage reduces the chemical potential and the interfacial electric field, thereby weakening the Rashba spin-orbit coupling and diminishing the strength of the SDE \cite{Yuan2022}. Meanwhile, increasing $B_y$ enhances the nonreciprocity in the junction, resulting in a monotonic increase in $\eta$ from $B_y$ = 0 to $B_y$ = 50 mT. The effect of $B_y$ on the SDE is further illustrated by plotting $\eta$ as a function of $B_y$ at fixed gate voltages. As shown in Fig.~\ref{fig1}~(f), $\eta$ increases with $B_y$ and reaches a maximum around $B_y$ = 50 mT. Beyond this point, further increases in $B_y$ result in a gradual decrease in $\eta$. This nonmonotonic behavior is consistent with both theoretical predictions and prior experimental observations \cite{Lotfizadeh2024}. The field at which the $\eta$ peaks provides insight into the strength of the spin-orbit coupling in the system \cite{Lotfizadeh2024}, as it reflects the optimal balance between spin-orbit and Zeeman energies that maximizes nonreciprocal transport. 

 In Fig.~\ref{fig2}~(a), we plot $\eta$ as a function of $B_y$ for various TG voltages. While the overall trend is similar for different TG voltages, we observe that the magnitude of $\eta$ changes with TG voltages. Traces obtained at TG below -1.6 V give a significant enhancement to $\eta$ compared to traces obtained at 0 V and -1 V. Notably, $\eta$ changes sign and becomes negative when TG = 0, 1 V and $B_y$ $>$ 100 mT. In Fig.~\ref{fig2}~(b), we present a bias current versus TG scan at $B_y$ = 40 mT to illustrate how the critical currents and the diode efficiency $\eta$ evolve with TG. The critical current first decreases as TG depletes the ends of the junction around -2.5 V and then remains nearly constant for more negative TG voltages (see SM for more gate dependence) while the positive branch is always larger than the negative branch. We repeat the TG vs current bias sweep at $B_y$ = 110 mT, as shown in Fig.~\ref{fig2}~(c). For TG = 0 V, the critical current at negative bias $I_{c-}$ exceeds the positive-bias critical current $I_{c+}$,  resulting in a negative diode efficiency. When TG is made more negative, $I_{c+}$ gradually increases while $I_{c-}$ decreases, This crossover leads to a sign change in $\eta$, which becomes positive for TG $<$ −2 V. Notably, the average critical current $\frac{I_{c+} + I_{c-}}{2}$ remains nearly constant throughout the TG sweep in Fig.~\ref{fig2}~(c), indicating that the chemical potential within the junction and the coupling between the junction and the superconducting leads are not significantly affected by TG. In Fig.~\ref{fig2}~(d) and (e), we plot absolute values of the positive- and negative-bias critical currents as functions of $B_y$ for two representative TG voltages. For TG = -0.6 V, both the $I_{c+}$ and the $I_{c-}$ decline with increasing $B_y$, and the difference between them peaks around $B_y$ = 50 mT, consistent with the typical behavior of the SDE observed in Fig.~\ref{fig2}~(a). For TG = -3 V, $I_{c+}$ increases significantly from 0 to 50 mT, while $I_{c-}$ remains nearly unchanged. This asymmetric evolution of the critical currents leads to a more pronounced enhancement of the diode efficiency $\eta$.

To further study the influence of TG on the diode efficiency $\eta$, we plot $\eta$ as a function of TG voltage in Fig.~\ref{fig2}~(f) for various $B_y$ at JG = 0. Notably, the dependence of  $\eta$ on TG exhibits a markedly different behavior compared to its dependence on JG. As the voltage on TG decreases, $\eta$ initially decreases and reaches a minimum near TG = -0.6 V. Upon further depletion, $\eta$ begins to increase and reaches a maximum around TG = -2.5 V. Interestingly, this voltage corresponds to the depletion threshold of the 2DEG in the etched hole regions (see Supplemental Material). The observed nonmonotonic dependence of $\eta$ on TG is counterintuitive and cannot be readily explained by changes in the chemical potential within the junction. This is because the JG screens the electric field from TG, thereby minimizing its influence on the potential in the center of the junction. Moreover, as shown in Fig.1(e), a decrease in chemical potential induced by JG depletion leads to a reduction in  $\eta$. This further supports the notion that the nonmonotonic dependence of $\eta$ on TG cannot be attributed to changes in the chemical potential within the junction. 
Thus, we suspect that the enhancement in $\eta$ at more negative TG voltages may arise from a novel and distinct mechanism. 

\begin{figure}[h]
    \centering
    \includegraphics[width=1\linewidth]{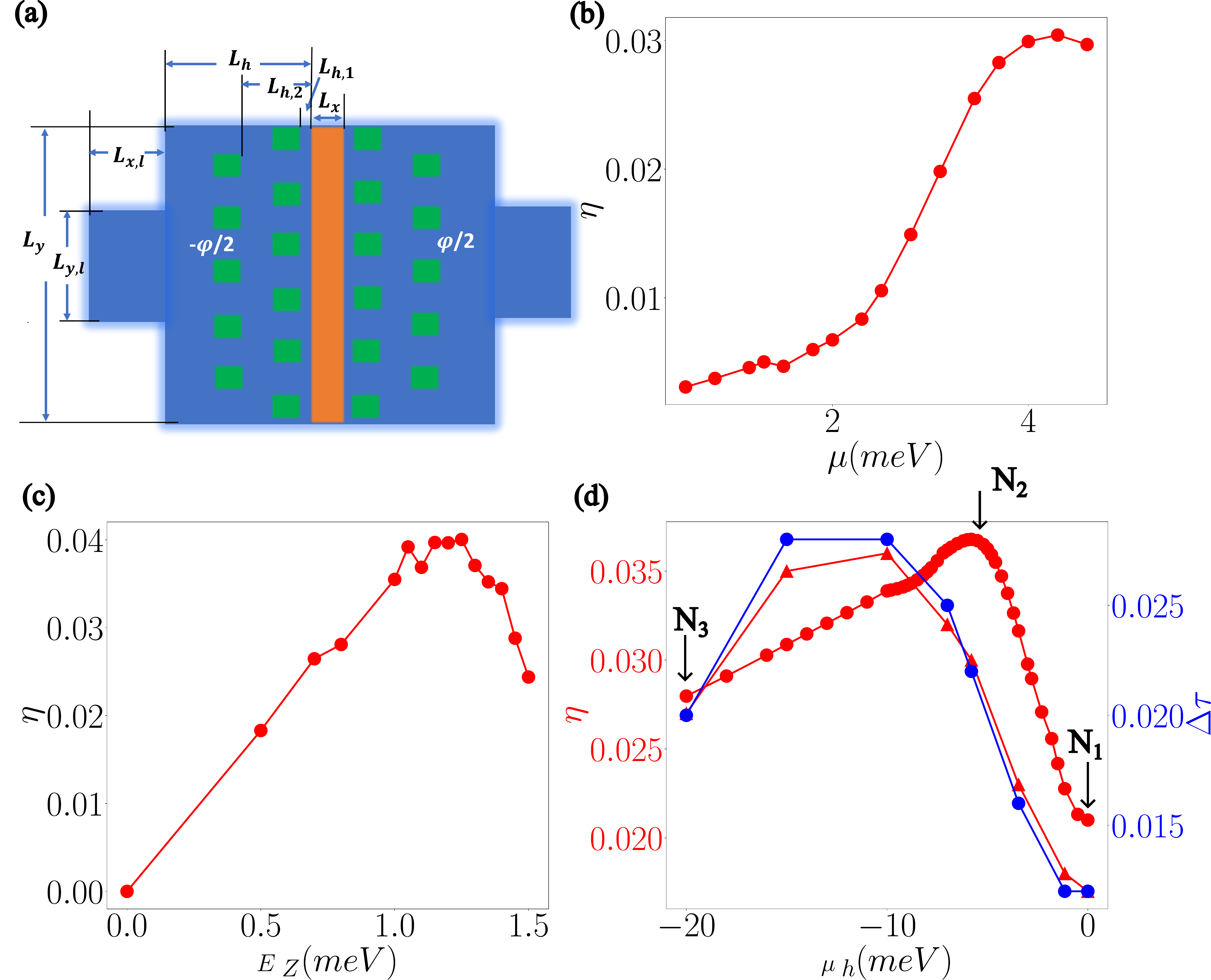}
    \caption{(a) Schematic of the punched hole setup in the simulations. The orange strip denotes the normal region; the blue region represents the superconducting area and the green squares represent the depleted holes. (b) Diode factor $\eta$ as a function of the normal region chemical potential $\mu$ at $E_Z=1$~meV. (c) Diode effect as a function of the Zeeman term. (d) Diode effect vs the chemical potential $\mu_h$ in the holes, denoted by the red circles. The fitted $\eta$ using the two-channel model is shown by the red triangles and $\Delta\tau=\tau_2-\tau_1$ is denoted by the blue circles. }
    \label{fig:hole}
\end{figure}

To understand the origin of the diode factor enhancement by depletion, we numerically simulated the punched-hole structure with rescaled parameters (see SM for details). Fig.~\ref{fig:hole}~(a) shows the geometry used for the simulations. Given the large width ($\sim 4\mu$m) of the experimental devices, to be able to complete the numerical calculations in reasonable times we rescaled the device's length, keeping, however, the ratios between all the physical lengths as close as possible to the experimental ones.
 
 As shown in Figs.~\ref{fig:hole}(b),~(c),~(d), the numerical results reproduce semiquantitatively the experimental results. Fig.~\ref{fig:hole}(b) shows the $\eta$'s dependence on the chemical potential $\mu$ (tuned by the JG's gate) inside the normal region. 
 Fig.~\ref{fig:hole}~(c), shows the dependence on the Zeeman energy, $E_Z$, Fig.~\ref{fig:hole}(c) for fixed $\mu$ and $\mu_h$. 
 The most compelling results are the ones of Fig.~\ref{fig:hole}~(d) 
 which shows the evolution of $\eta$ as $\mu_h$, i.e., the TG's gate, is varied for fixed $\mu$ and $E_Z$.

\begin{figure}[h]
    \centering
    \includegraphics[width=1\linewidth]{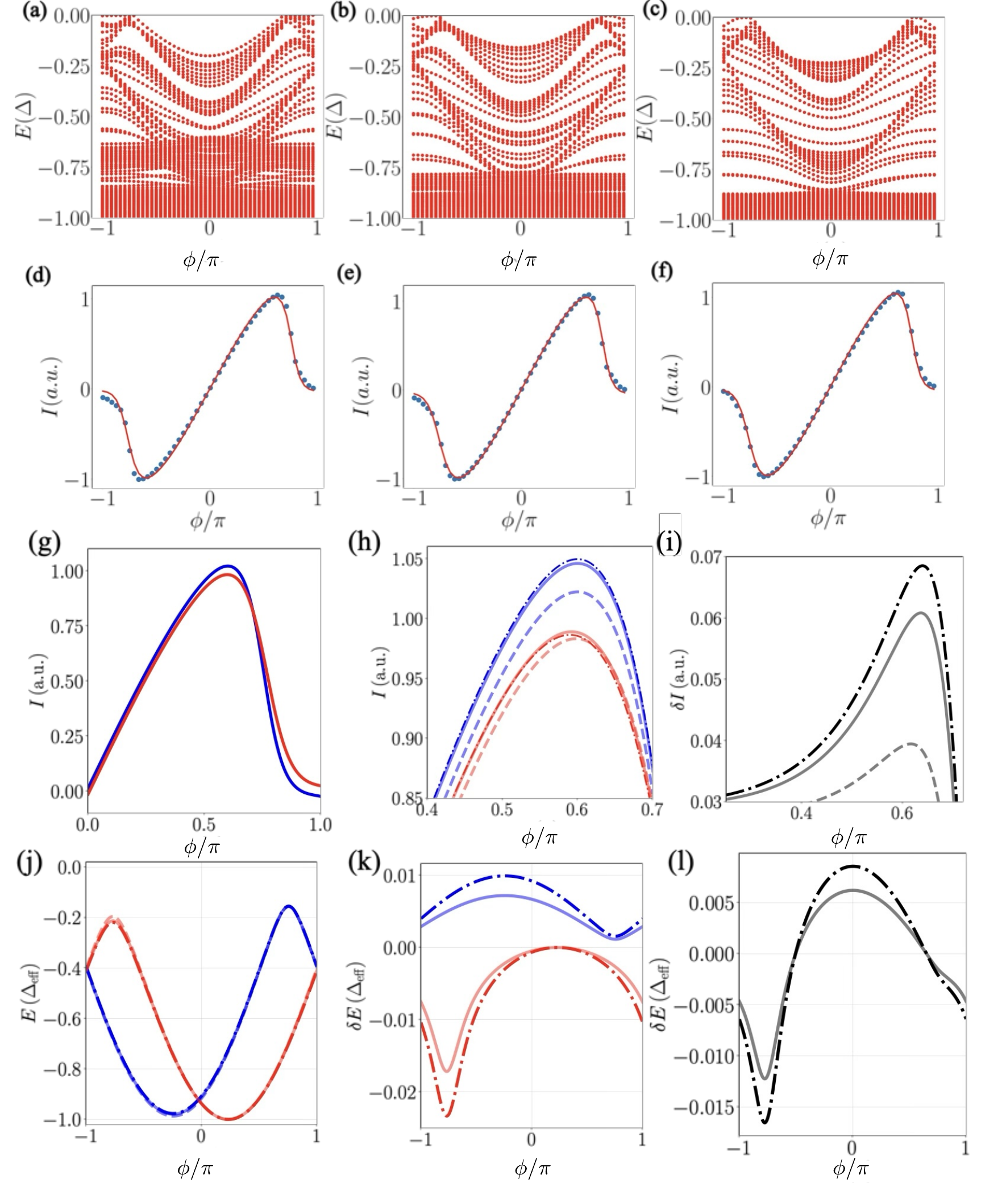}
    \caption{Spectrum at (a) $N_1$ point, (b) $N_2$, and (c) $N_3$. Corresponding CPR (dots) and two-channel fitting (line) is presented in (d)-(f). In the fitting, $\theta_1=0.77,
\Delta\theta=-1.5,\tau_1=0.975$ are fixed, while $I_1= 0.376,0.386,0.385$, and $\Delta\tau=-0.013,-0.023,-0.020$ in (d)-(f).
 Comparison of CPR at $\phi$ (blue) and $-\phi$ (red) where $\phi>0$, for (g) $N_1$ only, (h) $N_1$ (dashed lines), $N_2$ (dash-dotted lines), and $N_3$ (solid lines). (i). The difference of CPR at $\phi$ and $-\phi$ for  $N_1$ (dashed lines), $N_2$ (dash-dotted lines), and $N_3$ (solid lines). (j). Energy spectrum of the two effective modes (red for the first Andreev bound state and blue for the second) used in the fitting for $N_1$ (dashed lines), $N_2$ (dash-dotted lines), and $N_3$ (solid lines). (k). Energy difference between $N_1$ and $N_2$ (dash-dotted line) and between $N_1$ and $N_3$ (solid line), of the first mode (red) and the second mode (blue). (l). Difference between $N_1$ and $N_2$ (dash-dotted line) or $N_3$ (solid line), of the energy sum from the two modes.  }
    \label{fig:fitting}
\end{figure}

Given the large JJ's width, the supercurrent across the junction is due to the contribution of several Andreev bound states, shown in Figs.~\ref{fig:fitting}~(a)-(c). To try to understand the effect of TG qualitatively, we consider the simplest JJ's model that exhibits a diode effect: a model with just two modes having different transparencies, $\tau_1$, and $\tau_2=\tau_1+\Delta\tau$.  In this model, the dependence of the normalized supercurrent, $I=(\tilde I/I_1)$, on the phase $\phi$ across the JJ is given by the following equation:
\begin{equation}\label{eq:two-channel}
I=\frac{\sin(\phi+\theta_1)}{\sqrt{1-\tau_1\sin^2(\frac{\phi+\theta_1}{2})}}+\frac{\sin(\phi+\theta_1+\Delta\theta)}{\sqrt{1-(\tau_1+\Delta\tau)\sin^2(\frac{\phi+\theta_1+\Delta\theta}{2})}}.
\end{equation}
$I_i$ is just an overall scale that, for instance, depends on the JJ's width. The values of $\theta_1$, $\tau_1$ are obtained as those that best fit the numerical results for a given value of $\mu_h$, i.e., of the TG's gate, and are not varied as $\mu_h$ is tuned. This is physically well motivated and the parameters reduce to just $\Delta\tau$ and $\Delta\theta$ to model qualitatively the effect of $\mu_h$ on $\eta$.
%

Figures~\ref{fig:fitting}~(d)-(f) show, as dots, the current obtained using the full numerical model, i.e., summing up the conribuion of all the modes in the JJ, and the red lines show the best fits obtained using Eq.~\ceq{eq:two-channel}, the constraints on the fitting parameters discussed above, and the parameters' values provided in the figure caption, for the values of $\mu_h$ corresponding the the points $N_1$, $N_2$, $N_3$, in Fig.~\ref{fig:hole}~(d)
Similarly, Figs.~\ref{fig:fitting}~(g)-(i) compare the positive $\phi$ and negative $\phi$ current-phase relation to emphasize the presence of the SDE and its evolution with $\mu_h$.
Figures~\ref{fig:fitting}~(j)-(k) compare the dispersion of the two effective Andreev modes used to fit the CPR obtained fully numerical.

Results like the ones shown in Fig.~\ref{fig:fitting}~(d)-(l) allow us to obtain the evolution of $\eta$ with $\mu_h$. This is shown by the red triangles shown in Fig.~\ref{fig:hole}~(d). We can see that the simple two-modes model is able to capture, at least qualitatively, the non-monotonic dependence of $\eta$ on $\mu_h$. 
Expanding Eq.\eqref{eq:two-channel} up to second harmonics for $\eta$ we find the simple expression in terms of $\Delta\tau$:
\begin{equation}
\begin{aligned}\label{eq:final}
\eta\propto \Delta\tau\sin\frac{\Delta\theta}{2}
\end{aligned}
\end{equation}
The blue dots in Fig.~\ref{fig:hole}~(d) show $\eta$ obtained by also keeping fixed the 
value of the parameter $\Delta\theta$. These results demonstrate that the likely dominant mechanism for the non-monotonic scaling of $\eta$ with $\mu_h$ is the non-monotonic evolution of the most transparent effective Andreev bound state in the two-band model.

\begin{figure}
    \centering
    \includegraphics[width=1\linewidth]{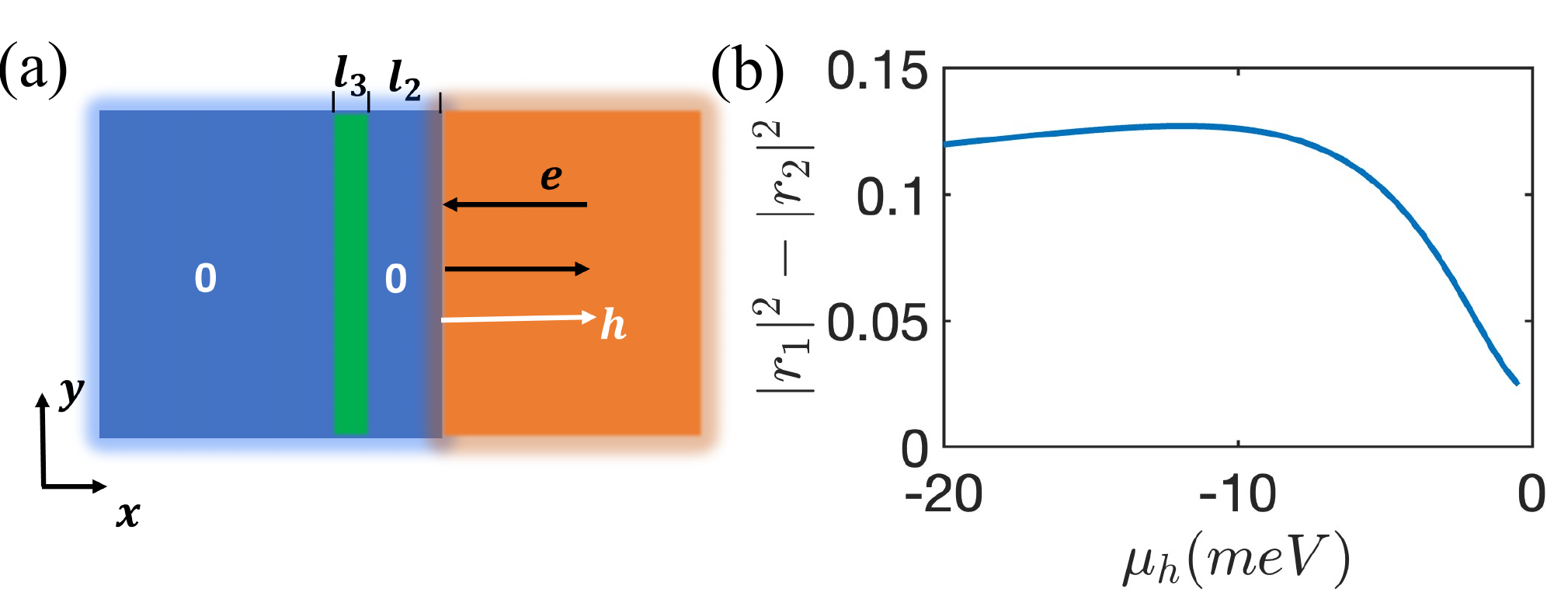}
    \caption{(a) Schematic of a simplified 1D-model for calculating the reflection. The color code is the same as Fig.~\ref{fig:hole}. The green region represents the depleted area without superconductivity, the orange area represents the semi-infinite normal region, and the blue denotes the superconducting area. The two separate superconducting areas have the same phase, and the left one is semi-infinite. (b) The difference in Andreev reflection coefficients between two channels differing only by their Fermi momentum. $r_1, r_2$ are the Andreev reflection coefficients of each channel, respectively.}
    \label{fig:reflect}
\end{figure}

To understand the evolution of $\Delta\tau$ as the depletion increases, we study a simplified 1D model, see Fig.~\ref{fig:reflect}(a). Since the transparency of Josephson junctions is the probability for Cooper pairs to cross the junction, in first approximation, it can be taken to be proportional o $|r_{eh}|^2=1-|r_{ee}|^2$ where $r_{eh}\, (r_{ee})$ represents the Andreev (normal) reflection coefficient \cite{beenakker1,beenakker2}. 
We then check the difference of the Andreev reflection coefficients between two states with different $k_F$ representing two effective channels, see Fig.~\ref{fig:reflect}(b). It is clear that the difference has a non-monotonic behavior as the depletion increases, with a peak at a certain optimal value of $\mu_h$. Importantly, this non-monotonic behavior of $\eta$ is independent of the distance between the depleted and normal regions, differently from the behavior of the supercurrent as discussed in Ref.~\cite{Yu2025} \footnote{See SM}. This shows that the difference of the Fermi momentum in the direction of the supercurrent 
results in different transparency. This difference non-monotonically changes with depletion, leading to the diode factor enhancement observed at certain depletion strengths.

In summary, we fabricated Josephson junctions with periodic hole structures on the Al contact
leads on InAs heterostructures. Through the depletion of the hole region by a top gate, we observe an enhanced superconducting diode effect, while the overall supercurrent is not significantly modified. As the effect of the top gate on the weak link region is screened by the junction gate, the enhancement is not induced by the change of chemical potential in the weak link region. Based on numerical simulations, we attribute the enhancement to the increased difference in transparency between different JJ's modes. Our observed enhancement differs from previous works, where a change in chemical potential or spin-orbit coupling induces the modification of the superconducting diode effect. As demonstrated by the unique tunability of our device, we believe such a periodic hole structure could be potentially useful in the development of novel superconducting devices.

\subsection
{Acknowledgments}
The work was funded by the US Department of Energy, Office of Basic Energy Sciences, via Award  DE-SC0022245. 
H.F. and E.R. thank Joseph J. Cuozzo for helpful discussions during the initial stages of the work.
Sandia National Laboratories is a multi-mission laboratory managed and operated by National Technology \& Engineering Solutions of Sandia, LLC (NTESS), a wholly owned subsidiary of Honeywell International Inc., for the U.S. Department of Energy’s National Nuclear Security Administration (DOE/NNSA) under contract DE-NA0003525. This written work is authored by an employee of NTESS. The employee, not NTESS, owns the right, title and interest in and to the written work and is responsible for its contents. Any subjective views or opinions that might be expressed in the written work do not necessarily represent the views of the U.S. Government. The publisher acknowledges that the U.S. Government retains a non-exclusive, paid-up, irrevocable, world-wide license to publish or reproduce the published form of this written work or allow others to do so, for U.S. Government purposes. The DOE will provide public access to results of federally sponsored research in accordance with the DOE Public Access Plan.

\bibliography{Ref.bib}

\end{document}